\documentclass[preprint]{appolb}
\usepackage{graphicx}
\usepackage{amsmath,amssymb,amsthm,cite, multirow}

\def\runhead{\relax}
\begin{document}
\title{Measurement of strong coupling constant by using event shape moments in perturbative theory%
}
\author{L. Khajooee, T. Kalalian, R. Saleh-Moghaddam, A. Sepehri, M. E. Zomorrodian
\address{Department of Physics, Faculty of sciences, Ferdowsi university of Mashhad, 91775-1436, Mashhad, Iran}
}
\maketitle
\begin{abstract}
We measure the strong coupling constant at \textit{NNLO} corrections. We do this analysis with moments of event shape variables: thrust, C parameter, heavy hemisphere mass, wide and total jet broadening, by fitting the \textit{L3} and \textit{DELPHI} data with \textit{NNLO} model. Our real data are consistent with \textit{NNLO} calculations, because it involves higher order terms in \textit{QCD} calculations.
\end{abstract}

\section{Introduction} 
Analyses of events originating from ${e}^{+} e{}^{- }$ annihilation into hadrons allow for studies \cite{1, 2, 3, 4, 5, 6} of Quantum Chromo Dynamics (\textit{QCD}), the theory of the strong interaction \cite{7, 8, 9, 10, 11, 12}. Comparison of observables such as jet production rates or event shapes with theoretical predictions provides access to the determination of the strong coupling ($\alpha _{s} $). Recently significant progress in the theoretical calculations of event shape moments and three jet rates has been made \cite{13}.
  
Event shape variables are interesting for studying the interplay between perturbative and non-perturbative dynamics. Apart from distributions of these observables one can study mean values and higher moments. The $n\ th$ moment of an event shape observable $Y$ is defined by \cite{14}:                
\begin{equation} \label{GrindEQ__1_} 
\left\langle y^{n} \right\rangle =\, \frac{1}{\sigma _{had} } \int _{\, \, \, 0}^{\, \, \, y_{\max } }y^{n}  \, \frac{d\sigma }{dy} dy 
\end{equation} 
  
where ${y}_{\max}$ is the kinematically allowed upper limit of the observable. 
  
Measurements of the strong coupling ($\alpha _{s} $) of \textit{QCD,} the theory of strong interaction, using different observables and different analysis methods serve as an important consistency test of \textit{QCD}. 
  
In section 2, we perform a new extraction of $\alpha _{s} $ from \textit{L3} and \textit{DELPHI} data with Next to Next Leading Order (\textit{NNLO}) model and measure the strong coupling constant at \textit{NNLO} then we compare the moment observables with other experiments. The last section includes our conclusions. 
  
\section{NNLO corrections to event shape moments in electron positron annihilation} 
  
\subsection{Definition of the observables}\label{ss1}
  
The properties of hadronic events may be characterized by a set of event shape observables. In this subsection \ref{ss1} briefly recall the definitions of the relevant event shape observables. The event shape variables are defined by the following sentences. 

\textbf{Thrust (T)} defined by the expression \cite{15, 16, 17, 18}: 
\begin{equation} \label{GrindEQ__2_} 
T=\max \left(\frac{\sum _{i}\left|\vec{P}_{i.} \vec{n}\right| }{\sum _{i}\left|p_{i} \right| } \right) 
\end{equation} 
  
The thrust axis $\vec{n}_{T} $ is the direction $\vec{n}$ which maximizes the expression in parentheses the value of the thrust can vary between $0.5$ and $1$. 
  
A plane through the origin and perpendicular to $\vec{n}_{T} $ divides the event into two hemispheres ${H}_{1}$ and ${H}_{2}$.

\textbf{C-Parameter:}   
The linearized momentum tensor $\theta ^{ij} $ is defined by \cite{19, 20}: 
  
\begin{equation}
\theta ^{ij} =\frac{1}{\sum _{1}\left|\vec{P}_{L} \right| } \sum _{k}\frac{P_{k}^{i} P_{k}^{i} }{\left|\vec{P}_{k} \right|}   ,                     i,j =1,2,3  
\end{equation}
Where the sum runs over all final state particles and ${P}_{k}^i $ is the $i\ th$ component of the three -- momentum $\vec{P}_{k} $of particle $k$ in the center of mass system. The tensor $\theta $ is normalized to have unit trace. In terms of the eigenvalues of the $\theta $ tensor, $\lambda _{1} ,\lambda _{2} ,\lambda _{3} ,with\lambda _{1} +\lambda _{2} +\lambda _{3} =1,$ one defines 
\begin{equation} \label{GrindEQ__4_} 
C=3(\lambda _{1} \lambda _{2} +\lambda _{2} \lambda _{3} +\lambda _{3} \lambda _{1} ) 
\end{equation} 
  
The c-parameter exhibits in perturbation theory a singularity at the three-parton boundary $c=\frac{3}{4} $

\textbf{Heavy hemisphere mass ( $\rho$):} The hemisphere masses are defined by \cite{21} 
\begin{equation} \label{GrindEQ__5_} 
M_{i}^{2} =\, \left(\begin{array}{l} {\sum Pj } \\ {j\in H_{i} } \end{array}\right)^{2}, i=1,2 
\end{equation} 
  
where ${P}_{j}$ denotes the four-momentum of particle j. The heavy hemisphere mass ${M}_{H}$ is then defined by: 
\begin{equation} \label{GrindEQ__6_} 
M_{H}^{2} =\, \max \, \left(M_{1}^{2} \, ,\, M_{2}^{2} \right) 
\end{equation} 
  
It is convenient to introduce the dimensionless quantity 
\begin{equation} \label{GrindEQ__7_} 
\rho =\frac{M_{H}^{2} }{Q^{2} }  
\end{equation} 
  
where $Q$ is the centre of mass energy. In leading order the distribution of the heavy hemisphere mass ($\rho $) is identical to the distribution of $(1-T)$.

\textbf{Jet broadening observables ${B}_{T}$ and ${B}_{W}$: } 
  
The hemisphere broadening \cite{22, 23, 24, 25, 26} are defined by
\begin{equation} \label{GrindEQ__8_} 
B_{K} =\, \left(\frac{\sum _{i\in \, H_{K} }\left|\vec{p}_{i} \times \vec{n}_{T} \right| }{2\sum _{i}\left|\vec{p}_{i} \right| } \right) 
\end{equation} 
  
For each of the two event hemispheres, ${H}_{K}$ defined above. The two observables are defined by:

\begin{equation} \label{GrindEQ__9_} 
B_{T} =B_{1} +B_{2} \, \, \, \, \, \, \, \, \, \, \, \, ,\, and\, \, \, \, \, \, B_{W} =\max \, \left(B_{1} ,B_{2} \right) 
\end{equation} 

where ${B}_{T}$ is the total and ${B}_{W}$ is the wide jet broadening. 
\subsection{Theoretical framework}  
The perturbative QCD expansion for the moment of the event shape observable $y$ up to NNLO at centre-of- mass energy $\sqrt{S} $ for renormalization scale $\mu ^{2} =s$ and $\alpha _{s} \equiv \, \alpha _{s} \left(S\right)$ is given by \cite{14}: 
\begin{equation} \label{GrindEQ__10_} 
\left\langle y^{n} \right\rangle \, \, \left(s,\, \mu ^{2} =s\right)=\, \left(\frac{\alpha _{s} }{2\pi } \right)^{1} \, \overline{A}_{y,n} +\, \left(\frac{\alpha _{s} }{2\pi } \right)^{2} \, \overline{B}_{y,n} +\, \left(\frac{\alpha _{s} }{2\pi } \right)^{3} \, \overline{C}_{y,n} +O\left(\alpha _{s} \right)^{4}  
\end{equation} 
  
The detailed calculations of the coefficients ${A}_{y,n}$ , ${B}_{y,n}$  and ${C}_{y,n}$ was achieved by Gehrmann et.al, \cite{14, 27}. In addition $\overline{A}_{y,n} ,\, \, \overline{B}_{y,n} \, \, and\, \, \, \overline{C}_{y,n} $ are related to ${A}_{y,n}$ , ${B}_{y,n}$  and ${C}_{y,n}$ according to;

\begin{align}\label{GrindEQ__11_} 
\overline{A}_{y,n} &=\, A_{y,n} \cr
\overline{B}_{y,n} &=\, B_{y,n} -\, \frac{3}{2} \, C_{F} \, \, A_{y,n}  ,\cr
\overline{C}_{y,n} &=\, C_{y,n} -\frac{3}{2} \, C_{F} B_{y,n} +\left(\frac{9}{4} C_{F}^{2} -K_{2} \right)A_{y,n}  
\end{align}

The constant ${K}_{2}$ is given by \cite{28, 29} 
\begin{equation} \label{GrindEQ__12_} 
K_{2} =\frac{1}{4} \left[-\frac{3}{2} C_{F}^{2} +C_{F} \, C_{A} \left(\frac{123}{2} -44\zeta _{3} \right)+C_{F} T_{R} N_{F} \left(-22+16\zeta _{3} \right)\right] 
\end{equation}

Where the QCD color factors are: 
\begin{equation} \label{GrindEQ__13_} 
C_{A} =N\, ,\, \, C_{F} =\frac{N^{2} -1}{2N} ,\, \, \, \, T_{R} =\frac{1}{2}  
\end{equation}

For $N=3$ colours and ${N}_{F}$ light quark flavors. The coefficients $A$, $B$ and $C$ have been computed for several event-shape variables In terms of the running coupling $\alpha _{s} \left(\mu ^{2} \right)$ the NNLO expression for an event shape moment measured at centre-of-mass energy squared s becomes \cite{14}:

\[\left\langle y^{n} \right\rangle \, \left(s,\mu ^{2} \right)=\, \left(\frac{\alpha _{s} \left(\mu \right)}{2\pi } \right)\, \overline{A}_{y,n} +\, \left(\frac{\alpha _{s} \left(\mu \right)}{2\pi } \right)^{2} \left(\overline{B}_{y,n} +\overline{A}_{y,n} \, \beta _{\circ } \, \log \, \frac{\mu ^{2} }{s} \right)+\left(\frac{\alpha _{s} \left(\mu \right)}{2\pi } \right)^{3} \] 
\begin{equation} \label{GrindEQ__14_} 
\left(\overline{C}_{y,n} +2\overline{B}_{y,n} \, \beta _{\circ } \, \log \, \frac{\mu ^{2} }{s} +\overline{A}_{y,n} \left(\beta _{\circ }^{2} \, \log ^{2} \, \frac{\mu ^{2} }{s} +\beta _{1} \, \log \, \frac{\mu ^{2} }{s} \right)\right)+o\left(\alpha _{s}^{4} \right). 
\end{equation} 
  
In which

\begin{equation} \label{GrindEQ__15_} 
\begin{array}{l} {\beta _{\circ } =\, \frac{11C_{A} -4T_{R} N_{F} }{6} \, \, } \\ {\beta _{1} =\, \frac{17C_{A}^{2} \, \, -10C_{A} \, T_{R} \, N_{F} \, -6C_{F} \, T_{R} \, N_{F} }{6} } \end{array} 
\end{equation}

\section{physics results} 
Moments of event shapes have been measured by various ${e}^{+} e^{-}$ collider experiments at centre-of- mass energies ranging from $40$ GeV to $207$ GeV \cite{30, 31}. 
  
To determine $\alpha _{s} $ at each energy point, the measured distributions are fitted in the ranges of energies. In this paper we use the experimental data for the five variables at $\left\langle \sqrt{s}\right\rangle =200.2\ GeV$. The scale uncertainly is obtained by repeating the fit for different values of the renormalization scale in the interval $0.5\sqrt{s}\le \mu \le 2\sqrt{s}$ \cite{32}. 
  
\begin{figure}[Hhtp]
\centering
\includegraphics[scale=.8]{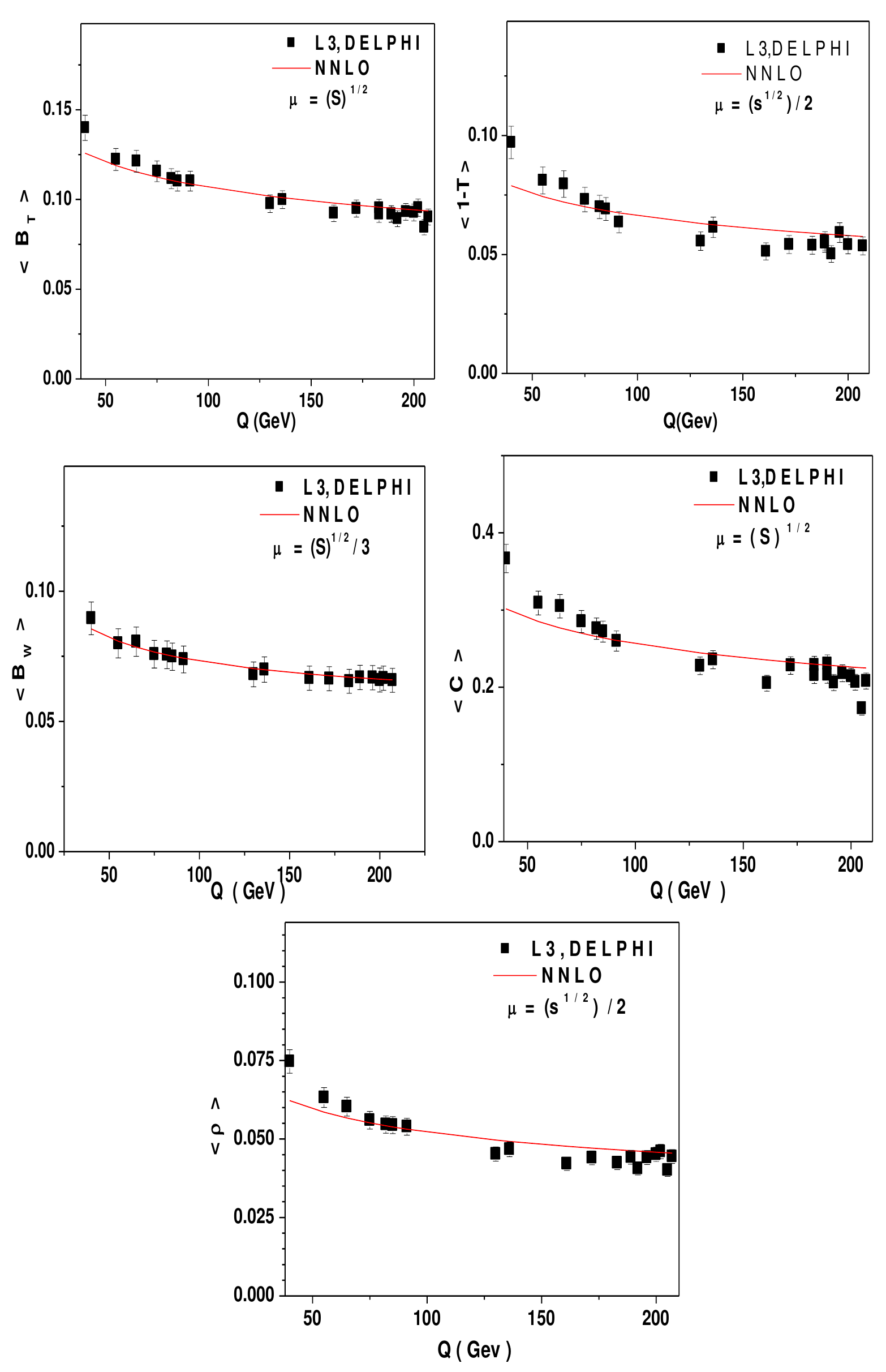}
\caption{First moments of Five event shape variables fitted with eq \eqref{GrindEQ__14_}, the data are from the $L3$ 
and DELPHI experiments, taken from refs \cite{30,31}. }\label{f1}
\end{figure}

\begin{figure}[Hhtp]
\centering
\includegraphics[scale=.8]{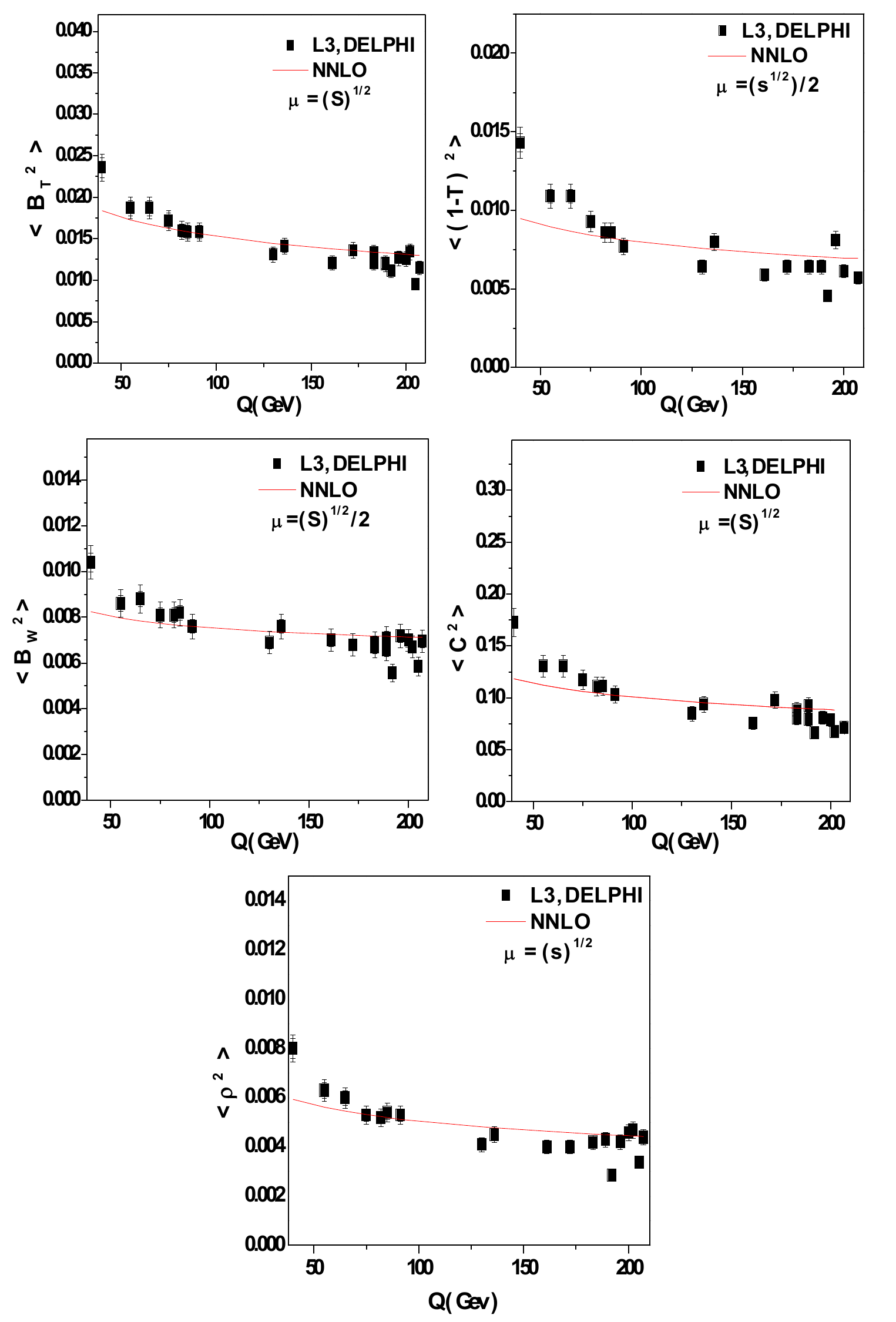}
\caption{Second  moments of Five event shape variables fitted with eq\eqref{GrindEQ__14_}, the data are from the L3 and DELPHI experiments, taken from refs \cite{30,31}. }\label{f2}
\end{figure}

The moments of the five standard event shapes are displayed in Figures \ref{f1} and \ref{f2}. The predictions are compared to ${L}_{3}$ and DELPHI data. Our real data are consistent with NNLO, compared with NLO or LO calculations, because it involves higher order terms in QCD calculations \cite{14}. 
  
\clearpage

The value of $\alpha _{s} $ was estimated by fitting the data \cite{30, 31} with NNLO expression for an event shape moment \eqref{GrindEQ__14_}. Separate fits were performed to each of the five observables at centre-of- mass energies ranging from $40$ GeV to $207$ GeV \cite{30, 31}. The fitted values for $\alpha _{s} $ change for different choices of the renormalization scale. This is demonstrated for the moments of event shapes in tables \ref{t1}-\ref{t5}. The given errors are the statistical errors. We also have indicated on each table the value of ${\alpha }_{{\rm S}}{\rm (}M_Z{\rm )}$ extracted at $M_{{\rm Z}}$ energy. We don't observe any significant change between the values of strong coupling constant for different event shape moments and renormalization scales.

\begin{table}[hbp]
\centering
\caption{Measurement of $\alpha_{s}$ from $\left\langle {(B}_T)\right\rangle $ and $\left\langle {{(B}_T)}^2\right\rangle $ moments for different choices of the renormalization scale.}\label{t1}
\begin{tabular}{|c|c|c|} \hline 
\textbf{Event shape variable} & $\mu $ & $\alpha _{s} $ \\ \hline
\multirow{5}{*}{$\left\langle {(B}_T)\right\rangle $} & ${\sqrt{s}}/{4}$ & $0.1327\pm 0.0008$ \\ \cline{2-3} 
 & $M_{{\rm Z}}$ & $0.1268\pm 0.0008$ \\ \cline{2-3} 
 & ${\sqrt{s}}/{3}$ & $0.1254\pm 0.0009$ \\ \cline{2-3} 
 & ${\sqrt{s}}/{2}$ & 0.1188$\pm 0.0009$ \\ \cline{2-3} 
 & $\sqrt{s}$ & $0.1131\pm 0.0009$ \\ \cline{2-3} 
 & $2\sqrt{s}$ & $0.1104 \pm 0.0009$ \\ \hline
\multirow{5}{*}{$\left\langle {{(B}_T)}^2\right\rangle $} & ${\sqrt{s}}/{4}$ & $0.1345 \pm 0.0022$ \\ \cline{2-3} 
 & $M_{{\rm Z}}$ & $0.1294\pm 0.0020$ \\ \cline{2-3} 
 & ${\sqrt{s}}/{3}$ & $0.1285\pm 0.0021$ \\ \cline{2-3} 
 & ${\sqrt{s}}/{2}$ & 0.1230$\pm$0.0021 \\ \cline{2-3} 
 & $\sqrt{s}$ & $0.1182\pm 0.0021$ \\ \cline{2-3} 
 & $2\sqrt{s}$ & $0.1159\pm 0.0020$ \\ \hline
\end{tabular} 
\end{table}

\begin{table}[H]
\centering
\caption{Measurement of $\alpha_{s}$ from $\left\langle 1-T\right\rangle $ and $\left\langle {\left(1-T\right)}^2\right\rangle $ moments for different choices of the renormalization scale.}\label{t2}
\begin{tabular}{|c|c|c|} \hline
\textbf{Event shape variable} & $\mu $ & $\alpha _{s} $ \\ \hline
\multirow{5}{*}{$\left\langle 1-T\right\rangle $} & ${\sqrt{s}}/{4}$ & $0.1339\pm 0.0024$ \\ \cline{2-3} 
 & $M_{{\rm Z}}$ & $0.1298\pm 0.0021$ \\ \cline{2-3} 
 & ${\sqrt{s}}/{3}$ & $0.1273\pm 0.0023$ \\ \cline{2-3} 
 & ${\sqrt{s}}/{2}$ & $0.1211\pm 0.0022$ \\ \cline{2-3} 
 & $\sqrt{s}$ & $0.1158\pm 0.0021$ \\ \cline{2-3} 
 & $2\sqrt{s}$ & $0.1133\pm 0.0021$ \\ \hline
\multirow{5}{*}{$\left\langle {\left(1-T\right)}^2\right\rangle $} & ${\sqrt{s}}/{4}$ & $0.1377\pm 0.0049$ \\ \cline{2-3}
 & $M_{{\rm Z}}$ & $0.1364\pm0.0039$ \\ \cline{2-3} 
 & ${\sqrt{s}}/{3}$ & $0.1316\pm0.0047$ \\ \cline{2-3} 
 & ${\sqrt{s}}/{2}$ & $0.1260\pm0.0045$ \\ \cline{2-3} 
 & $\sqrt{s}$ & $0.1210\pm0.0043$ \\ \cline{2-3} 
 & $2\sqrt{s}$ & $0.1186\pm0.0042$ \\ \hline
\end{tabular} 
\end{table}

\begin{table}[H]
\centering
\caption{Measurement of $\alpha{}_{s}$ from $\left\langle {(B}_W)\right\rangle $ and $\left\langle {{(B}_W)}^2\right\rangle$ moments for different choices of the renormalization scale.}\label{t3}
\begin{tabular}{|c|c|c|} \hline
\textbf{Event shape variable} & $\mu $ & $\alpha _{s} $ \\ \hline
\multirow{5}{*}{$\left\langle {(B}_W)\right\rangle $} & ${\sqrt{s}}/{4}$ & $0.1297\pm0.0009$ \\ \cline{2-3} 
 & $M_{{\rm Z}}$ & $0.1249\pm0.0006$ \\ \cline{2-3} 
 & ${\sqrt{s}}/{3}$ & $0.1228\pm0.0008$ \\ \cline{2-3} 
 & ${\sqrt{s}}/{2}$ & $0.1159\pm0.0007$ \\ \cline{2-3} 
 & $\sqrt{s}$ & $0.1097\pm0.0006$ \\ \cline{2-3} 
 & $2\sqrt{s}$ & $0.1068\pm0.0006$ \\ \hline
\multirow{5}{*}{$\left\langle {{(B}_W)}^2\right\rangle $} & ${\sqrt{s}}/{4}$ & $0.1231\pm0.0016$ \\ \cline{2-3} 
 & $M_{{\rm Z}}$ & $0.1195\pm0.0013$ \\ \cline{2-3} 
 & ${\sqrt{s}}/{3}$ & $0.1166\pm0.0015$ \\ \cline{2-3} 
 & ${\sqrt{s}}/{2}$ & $0.1107\pm0.0014$ \\ \cline{2-3} 
 & $\sqrt{s}$ & $0.1055\pm0.0014$ \\ \cline{2-3} 
 & $2\sqrt{s}$ & $0.1031\pm0.0013$ \\ \hline
\end{tabular}
\end{table}

\begin{table}[H]
\centering
\caption{Measurement of $\alpha _{s}$ from $\left\langle \ C\ \right\rangle $ and $\left\langle \ C^2\ \right\rangle $ moments for different choices of the renormalization scale.}\label{t4}
\begin{tabular}{|c|c|c|} \hline
\textbf{Event shape variable} & $\mu $ & $\alpha _{s} $ \\ \hline
\multirow{5}{*}{$\left\langle \ C\ \right\rangle $} & ${\sqrt{s}}/{4}$ & $0.1324\pm0.0019$ \\ \cline{2-3} 
 & $M_{{\rm Z}}$ & $0.1272\pm0.0017$ \\ \cline{2-3} 
 & ${\sqrt{s}}/{3}$ & $0.1258\pm0.0019$ \\ \cline{2-3} 
 & ${\sqrt{s}}/{2}$ & $0.1197\pm0.0018$ \\ \cline{2-3} 
 & $\sqrt{s}$ & $0.1144\pm0.0018$ \\ \cline{2-3} 
 & $2\sqrt{s}$ & $0.1120\pm0.0017$ \\ \hline 
\multirow{5}{*}{$\left\langle \ C^2\ \right\rangle $} & ${\sqrt{s}}/{4}$ & $0.1364\pm0.0039$ \\ \cline{2-3} 
 & $M_{{\rm Z}}$ & $0.1335\pm0.0029$ \\ \cline{2-3} 
 & ${\sqrt{s}}/{3}$ & $0.1305\pm0.0038$ \\ \cline{2-3} 
 & ${\sqrt{s}}/{2}$ & $0.1249\pm0.0036$ \\ \cline{2-3} 
 & $\sqrt{s}$ & $0.1200\pm0.0035$ \\ \cline{2-3} 
 & $2\sqrt{s}$ & $0.1177\pm0.0035$ \\ \hline 
\end{tabular} 
\end{table}
  
\begin{table}[H]
\centering
\caption{Measurement of $\alpha_{s}$ from $\left\langle \ \rho \ \right\rangle $ and $\left\langle \ {\rho }^2\ \right\rangle $  moments of the renormalization scale.}\label{t5}
\begin{tabular}{|c|c|c|} \hline
\textbf{Event shape variable} & $\mu $ & $\alpha _{s} $ \\ \hline
\multirow{5}{*}{$\left\langle \ \rho \ \right\rangle $} & ${\sqrt{s}}/{4}$ & $0.1367\pm0.0021$ \\ \cline{2-3} 
 & $M_{{\rm Z}}$ & $0.1304\pm0.0020$ \\ \cline{2-3} 
 & ${\sqrt{s}}/{3}$ & $0.1287\pm0.0020$ \\ \cline{2-3} 
 & ${\sqrt{s}}/{2}$ & $0.1215\pm0.0019$ \\ \cline{2-3} 
 & $\sqrt{s}$ & $0.1153\pm0.0018$ \\ \cline{2-3} 
 & $2\sqrt{s}$ & $0.1124\pm0.0018$ \\ \hline
\multirow{5}{*}{$\left\langle \ {\rho }^2\ \right\rangle $} & ${\sqrt{s}}/{4}$ & $0.1317\pm0.0035$ \\ \cline{2-3} 
 & $M_{{\rm Z}}$ & $0.1260\pm0.0033$ \\ \cline{2-3} 
 & ${\sqrt{s}}/{3}$ & $0.1246\pm0.0033$ \\ \cline{2-3} 
 & ${\sqrt{s}}/{2}$ & $0.1180\pm0.0030$ \\ \cline{2-3} 
 & $\sqrt{s}$ & $0.1124\pm0.0029$ \\ \cline{2-3} 
 & $2\sqrt{s}$ & $0.1098\pm0.0028$ \\ \hline
\end{tabular} 
\end{table}

\clearpage
  
We observe that our obtained values for coupling constant considering the NNLO corrections for different event shape variables are in good agreement with the other experiments \cite{10}. 

\section{Conclusions}
In this paper we have presented measurements of the strong coupling constant for hadronic events produced at $L3$ and DELPHI in the centre-of-mass energies $40$ GeV to $207$GeV. We measured the coupling constant considering the moments of the event shape observables using next-to-next leading order (NNLO) calculations for different choices of the renormalization scale. Our results are consistent with those obtained from other experiments, as well as with the QCD theory.

\end{document}